\pdfoutput=1
\documentclass[a4paper]{llncs}[2018/03/10]
\usepackage{etoolbox}
\makeatletter
\let\llncs@addcontentsline\addcontentsline
\patchcmd{\maketitle}{\addcontentsline}{\llncs@addcontentsline}{}{}
\patchcmd{\maketitle}{\addcontentsline}{\llncs@addcontentsline}{}{}
\patchcmd{\maketitle}{\addcontentsline}{\llncs@addcontentsline}{}{}
\makeatother
\usepackage{mathptmx}
\usepackage{helvet}
\usepackage{courier}
\usepackage{type1cm}
\usepackage{makeidx}
\usepackage[rgb,dvipsnames,x11names]{xcolor}
\usepackage{graphicx}
\usepackage{hyperref}
\usepackage[all]{hypcap}
\usepackage{bookmark}
\usepackage{cite}
\usepackage{amsfonts}
\usepackage[cmex10]{amsmath}
\usepackage{amssymb}
\usepackage{filecontents}

\DeclareGraphicsExtensions{.mps}
\hypersetup{%
	pdfdisplaydoctitle=true,
	pdftitle={On the Metropolis Algorithm for Urban Street Networks
		},
	pdfauthor={%
		J\'er\^ome~Benoit (ORCID: 0000-0003-1226-6757)
		and
		Saif~Eddin~Jabari (ORCID: 0000-0002-2314-5312)
		},
	pdfsubject={ComplexNetworks 2019
		- The 8th International Conference on Complex Networks and Their Applications
		- 10-12 December, 2019, Lisbon (Portugal)%
		},
	pdfkeywords={
		extended abstract
		},
	pdfcreator={\LaTeXe{} and its friends},
	pdfpagelayout=SinglePage,
	pdfpagemode=UseOutlines,
	pdfstartpage=1,
	pdfhighlight=/O,
	pdfview=FitH,
	pdfstartview=FitH,
	colorlinks=true,
	allcolors=RedOrange,
	citecolor=RoyalBlue3,
	urlcolor=RoyalBlue3,
	linkcolor=RoyalBlue3,
	bookmarksnumbered=true,
	bookmarksopen=true,
	bookmarksopenlevel=2,
	}

\begin{document}

\title{On the Metropolis Algorithm for Urban Street Networks}
\titlerunning{On the Metropolis Algorithm for Urban Street Networks}
\author{%
	J\'er\^ome Benoit\inst{1}
	\and
	Saif~Eddin Jabari\inst{1,2}
	}
\authorrunning{J\'er\^ome Benoit and Saif~Eddin Jabari}
\tocauthor{J\'er\^ome Benoit and Saif~Eddin Jabari}
\institute{%
	New~York~University Abu~Dhabi,
	Saadiyat~Island,
	POB~129188,
	Abu~Dhabi,
	UAE\\
	\and
	New York University Tandon~School~of~Engineering,
	Brooklyn,
	NY 11201,
	New~York,
	USA\\
	\email{jerome.benoit@nyu.edu}
	}
\maketitle

\section{Introduction}

The complexity of urban street networks
is well accepted to reside
in the \emph{information space} where
roads map to nodes and
junctions map to links~\cite{MRosvall2005,BJiangSZhaoJYin2008,JBenoitOPSOUSN2019,SESOPLUSN}.
By investigating the information space,
we aim to provide
new tools
to study our cities.

Striking
broad valence distributions
have been observed among actual information networks.
The urban street networks of self-organized cities
deserve special attention
for at least two reasons.
First,
they might have reached,
over time,
a spontaneous equilibrium
that most designed cities fail to reproduce~\cite{CAlexanderACINAT1965}.
Second,
the valence distributions of their information networks
clearly distinguish themselves
by following the (scale-free) Pareto distribution~\cite{BJiangSZhaoJYin2008,JBenoitOPSOUSN2019,SESOPLUSN}.
The second reason,
while it supports the first one,
may hopefully lead to a tractable theory.

Accordingly,
we envision urban street networks
as evolving social systems subject to
a Boltzmann-mesoscopic entropy preservation~\cite{JBenoitOPSOUSN2019,SESOPLUSN}.
This preservation ensures the passage from the road-junction hierarchy
to a scale-free coherence,
{i.e.},
that the valence distribution of the information network
follows a Pareto distribution.
The Boltzmann-mesoscopic entropies reflect
the perception that inhabitants have of their own city,
so they are better expressed in terms of \emph{surprisal}.
In brief,
we conjecture that
information networks tend to evolve
by maintaining their amount of surprisal constant on average.

Even so information networks are well recognized as relevant,
there is some art in how social and geographical criteria are pondered
to construct them from urban street networks.
However,
the \emph{deflection angles} between pairs of adjacent street-segments
appear to be pertinent constructing parameters~\cite{BJiangSZhaoJYin2008,JBenoitOPSOUSN2019,SESOPLUSN}.
Naive \emph{behavioural based join principles}~\cite{BJiangSZhaoJYin2008}
based on deflection angles
have been used with good success~\cite{BJiangSZhaoJYin2008}.
Amazingly,
the most successful one~\cite{BJiangSZhaoJYin2008} is a random process with numerous outputs.
The output arbitrariness must be addressed.
Embracing the idea that information networks
are driven by surprisal
allows us to elevate these principles
to a single-flip \emph{Metropolis algorithm}
as used for generating equilibrium states
of Ising-like models in statistical physics~\cite{MCMSP}.

\section{Method}

Here nodes are \emph{natural roads} (or roads for short),
that is,
an exclusive sequence of successive street-segments
joined according to some {behavioural based join principle}~\cite{BJiangSZhaoJYin2008}.
If beyond some threshold angle any joining has to be forbidden, multiple possibilities remain open.
Two join principles based on deflection angle have appeared realistic
against well-founded cadasters.
The \emph{self-best-fit} and
\emph{self[-random]-fit}
join principles
operate sequentially on growing roads,
until applicable,
by randomly seeding them with a not-yet-selected street-segment
before recursively appending,
until applicable,
one of the not-yet-appended street-segments.
The {self} join principles differ only by the choice of the nominees.
The {self-best-fit} join principle
picks the not-yet-appended street-segments whose deflection angle is the smallest.
By contrast,
the {self[-random]-fit} join principle
chooses randomly.
The {random} variant generally gives the best \textit{``fit''}~\cite{BJiangSZhaoJYin2008}.

Given a Boltzmann-mesoscopic system,
let us denote by $\Pr(\Omega)$
the distribution of the numbers of configurations $\Omega$ of its mesoscopic objects $o$.
If the average amount of surprisal
\begin{math}
	\sum_{\Omega}\Pr(\Omega)\ln\Omega
\end{math}
is preserved,
$\Pr(\Omega)$ most plausibly follows a Pareto distribution
\begin{math}
	\Pr(\Omega)\propto\Omega^{-\lambda}
\end{math}
by virtue of Jaynes's Maximum Entropy principle~\cite{JBenoitOPSOUSN2019,SESOPLUSN}.
Assuming that an information network is such a system,
the probability $p_{\mu}$ of its state $\mu$ yields
\begin{equation}\label{OMAUSN/eq/SOUSN/surprisal/probability}
	p_{\mu}\propto\prod_{o_{\mu}\in\{r_{\mu},j_{\mu}\}}\Omega_{o_{\mu}}^{-\lambda}=\mathrm{e}^{-\lambda{S_{\mu}}}
	\quad\text{with}\quad
	S_{\mu}=\sum_{o_{\mu}\in\{r_{\mu},j_{\mu}\}}\ln\Omega_{o_{\mu}}
\end{equation}
the amount of surprisal in state $\mu$;
the product (so the sum) runs over the roads $r_{\mu}$ and junctions $j_{\mu}$ of state~$\mu$.
This state probability depends only on the actual state of the information network.
So,
for a given self-organized urban street network,
we can generate \emph{Markov chains} \cite{MCMSP} of information networks
whose valence distribution reaches a Pareto distribution as equilibrium.

\begin{figure}[!bh]
	\begin{center}
		\includegraphics[width=\linewidth]{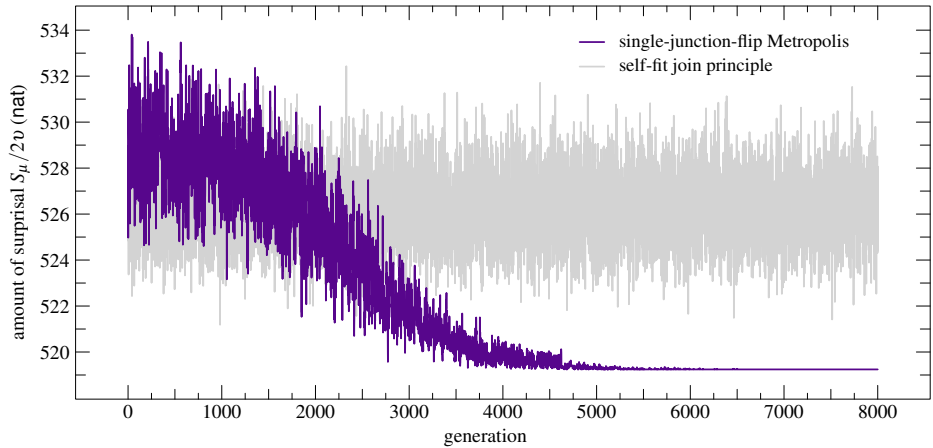}
	\end{center}
	\caption{\label{OMAUSN/fig/SOUSN/surprisal/Ahmedabad}%
		Typical single-junction-flip Metropolis run for Old Ahmedabad (India):
		the foreground purple
		generation series
		plots a typical run starting from a self-fit state;
		the background light-grey
		generation series
		plots
		a typical random sequence of self-fit states
		with the same \textit{modus operandi}.
		The simulations were run with a custom
		\texttt{\textbf{C}}
		code
		adopting and adapting
		typical techniques~\cite{MCMSP}.
		}
\vspace{-0.4cm}
\end{figure}

To achieve this, we must place on two conditions~\cite{MCMSP}.
First,
the \emph{condition of ergodicity} assures that
the Markov process can reach any state from any other one.
The {self-fit} join principle
readily inspires us the following \emph{single-junction-flip} ergodic iteration:
choose randomly a street-segment,
then a direction towards one of its junctions,
then a new street-segment nominee,
and
finally recompose accordingly the arrangement of the chosen junction.
Second,
the \emph{condition of detailed balance} assures both that
every Markov chain comes to an equilibrium
and that
it is the probability distribution \eqref{OMAUSN/eq/SOUSN/surprisal/probability}
which is effectively generated.
An abundant literature exists on the subject~\cite{MCMSP}.
For the sake of preliminary investigation,
we adopt the Metropolis algorithm~\cite{MCMSP}.
Thusly our \emph{acceptance ratio} $A(\mu\to\nu)$ to accept
a new state $\nu$ from state $\mu$ writes
\begin{equation}
	A(\mu\to\nu)=
		\begin{cases}
			\mathrm{e}^{-\lambda {(S_{\nu}-S_{\mu})}} & \text{if } {S_{\nu}-S_{\mu}}>0\\
			1 & \text{otherwise}.
		\end{cases}
\end{equation}
In fact,
we adapt the \emph{single-spin-flip} variant of the Metropolis algorithm~\cite{MCMSP}
since we use the {single-junction-flip} dynamics to generate new states.
We refer to the literature to elaborate more sophisticated variants~\cite{MCMSP}.

For early investigations,
we may reduce urban street networks to their roads only
so that the amount of surprisal $S_{\mu}$ in state $\mu$
simplifies~\cite{JBenoitOPSOUSN2019,SESOPLUSN} to
\begin{equation}
	S_{\mu}=2\upsilon\sum_{r_{\mu}}\ln{n_{r_{\mu}}}
	\quad\text{since}\quad
	\Omega_{r_{\mu}}\propto{n_{r_{\mu}}^{2\upsilon}}
\end{equation}
where $\upsilon$ is the \emph{number of vital connections}
for roads~\cite{JBenoitOPSOUSN2019,SESOPLUSN};
the sum runs over the roads $r_{\mu}$ of state~$\mu$.
We denote by $n_{r}$ the number of junctions of road $r$.

\section{Results and Discussion}

Figure~\ref{OMAUSN/fig/SOUSN/surprisal/Ahmedabad} shows
a typical single-junction-flip Metropolis run
for the urban street network of Old Ahmedabad.
Besides validating our approach,
our runs give two precious indications.
First,
the actual convergence of typical runs confirms that
information networks
of Old Ahmedabad
plausibly follow a scale-free coherence.
Second,
the convergent information network
of least surprisal
appear mostly unreachable
through the self-fit join principle.
Both encourage to reinforce
the similitude with Ising-like models~\cite{MCMSP}.

In future works,
we expect to generate among self-organized information networks different surprisal equilibria.
Ultimately,
this may bring us
a thermodynamic-like toolbox~\cite{MCMSP}
to investigate and understand
the geometrical rules and the social dynamics
that actually govern
urban street networks and,
by extension,
cities.


\end{document}